\documentclass[pra,twocolumn,
superscriptaddress,
longbibliography,
aps]{revtex4-2}
\usepackage{graphicx}
\usepackage{amsmath}
\usepackage{amsfonts}
\usepackage{amssymb}
\usepackage{mathptmx}
\usepackage[colorlinks=true,linkcolor=blue,urlcolor=blue,citecolor=blue]{hyperref}
\usepackage{color}

\newcommand{\beq}{\begin{equation}}
	\newcommand{\eeq}{\end{equation}}

\usepackage{dsfont}
\usepackage{placeins} 
\usepackage{bm}           

\usepackage{scalerel}
\usepackage{tikz}
\usetikzlibrary{svg.path}
\definecolor{orcidlogocol}{HTML}{A6CE39}
\tikzset{
	orcidlogo/.pic={
		\fill[orcidlogocol] svg{M256,128c0,70.7-57.3,128-128,128C57.3,256,0,198.7,0,128C0,57.3,57.3,0,128,0C198.7,0,256,57.3,256,128z};
		\fill[white] svg{M86.3,186.2H70.9V79.1h15.4v48.4V186.2z}
		svg{M108.9,79.1h41.6c39.6,0,57,28.3,57,53.6c0,27.5-21.5,53.6-56.8,53.6h-41.8V79.1z M124.3,172.4h24.5c34.9,0,42.9-26.5,42.9-39.7c0-21.5-13.7-39.7-43.7-39.7h-23.7V172.4z}
		svg{M88.7,56.8c0,5.5-4.5,10.1-10.1,10.1c-5.6,0-10.1-4.6-10.1-10.1c0-5.6,4.5-10.1,10.1-10.1C84.2,46.7,88.7,51.3,88.7,56.8z};
	}
}
\newcommand\orcid[1]{\href{https://orcid.org/#1}{\mbox{\scalerel*{
				\begin{tikzpicture}[yscale=-1,transform shape]
					\pic{orcidlogo};
				\end{tikzpicture}
			}{R}}}}

\begin{document}
\title{Cooling strongly self-organized particles using adiabatic demagnetization}
\author{Simon~B.~J\"ager\,\orcid{0000-0002-2585-5246}}
	\affiliation{Physikalisches Institut, University of Bonn, Nussallee 12, 53115 Bonn, Germany}
	\begin{abstract}
We study the dynamics of polarizable particles coupled to a lossy cavity mode that are transversally driven by a laser.  Our analysis is performed in the regime where the cavity linewidth exceeds the recoil frequency by several orders of magnitude. Using a two-stage cooling protocol we show that the particles' kinetic energy can be reduced down to the recoil energy. This cooling protocol relies in its first stage on a high laser power such that the particles cool into a strongly self-organized pattern. This can be seen as a strongly magnetized state. In a second stage we adiabatically ramp down the laser intensity such that the particles' kinetic energy is transferred to their potential energy and the particles are ``demagnetized''. In this second stage we optimize the ramping speed which needs to be fast enough to avoid unwanted heating and slow enough such that the dynamics remains to good approximation adiabatic. 
	\end{abstract}

	\maketitle
	
\section{Introduction}   
The realization of quantum technologies~\cite{Ladd:2010,Acin:2018,Barzanjeh:2022} based on polarizable particles such as atoms, ions, molecules, and nanoparticles relies on the precise control of their motional degrees of freedom. One important step to achieve full control of the particles is to reduce their residual motion. A key technique to achieve this is laser cooling \cite{Wineland:1979,Chu:1998,Wieman:1999,Tannoudji:1998,Phillips:1998,Stenholm:1986,Metcalf:1999} which can be used to achieve temperatures that leave the particles close to their zero-point motion. The basic principle behind laser cooling is the enhanced absorption rate of laser photons that lower the momentum of the particle. Subsequently, incoherent scattering of a photon from the particle into free space results in a lower kinetic energy of the particle. Despite the big success of laser cooling one major problem is that it typically relies on closed transitions and the atomic species at hand. This hinders the universal application of conventional laser cooling techniques to more complex systems like molecules or nanoparticles. 

A good candidate to overcome this problem is cavity cooling where the particles' motion is cooled by coherent scattering of laser photons \cite{Horak:1997,Vuletic:2000,Domokos:2001,Domokos:2002, Black:2003, Black:2003,Maunz:2004,Morigi:2007,Schleier-Smith:2011,Wolke:2012,Hosseini:2017}. 
Here, the particles' kinetic energy is carried away by the scattered cavity photons while the internal state of the particles' remain almost unaltered. The cavity mode is tuned in a way such that photons with larger energy than the drive are more likely to leave the cavity and can therefore not be reabsorbed by the particles. Consequently, the particles are, in average, at a lower kinetic energy than before the scattering event. In such a setup the minimum temperature is typically bounded by the linewidth of the cavity \cite{Domokos:2001}. Cavity cooling of single atoms~\cite{Maunz:2004} and collective cooling~\cite{Black:2003,Hosseini:2017} have been realized in experimental labs. Since cavity cooling does not rely on incoherent scattering from a specific internal state it is has been proposed for cooling molecules~\cite{Morigi:2007} and experimentally realized for cooling nanoparticles~\cite{Asenbaum:2013,Delic:2019}. Although, sub-recoil cooling has been achieved experimentally~\cite{Wolke:2012} in an optical cavity with very narrow linewidth usually the limit set by the cavity linewidth lies well above the recoil limit.

In this paper we investigate the situation when the cavity linewidth is orders of magnitude larger than the recoil frequency which is for instance the case for the experiment described in Ref.~\cite{Hosseini:2017} but also in several other experimental labs. We demonstrate that one can theoretically still achieve temperatures that are of the order of a single recoil by using a combination of cavity cooling and adiabatic control of optomechanical forces. The key ingredient is that the scattered photons besides cooling also mediate collective interactions which allow the particles to self-organize\cite{Domokos:2002,Asboth:2005}. Self-organization occurs if the driving-laser power exceeds a threshold determined by the cavity parameters and the temperature of the particles. Here, the particles form spontaneously a pattern with a spacing that is determined by the wavelength of the light and allows for constructive interference of scattered photons. Atomic self-organization has been observed with ultra-cold bosons~\cite{Baumann:2010}, thermal atoms~\cite{Arnold:2012} and ultra-cold fermions~\cite{Wu:2023,Helson:2023}. The formation of a self-organized pattern can be described as a ferromagnetic phase of a long-range interacting system where the collectively scattered light field can be understood as order parameter that measures the magnetization of the atomic ensemble\cite{Schuetz:2015}.

\begin{figure}[h!] 
	\center
	\includegraphics[width=1\linewidth]{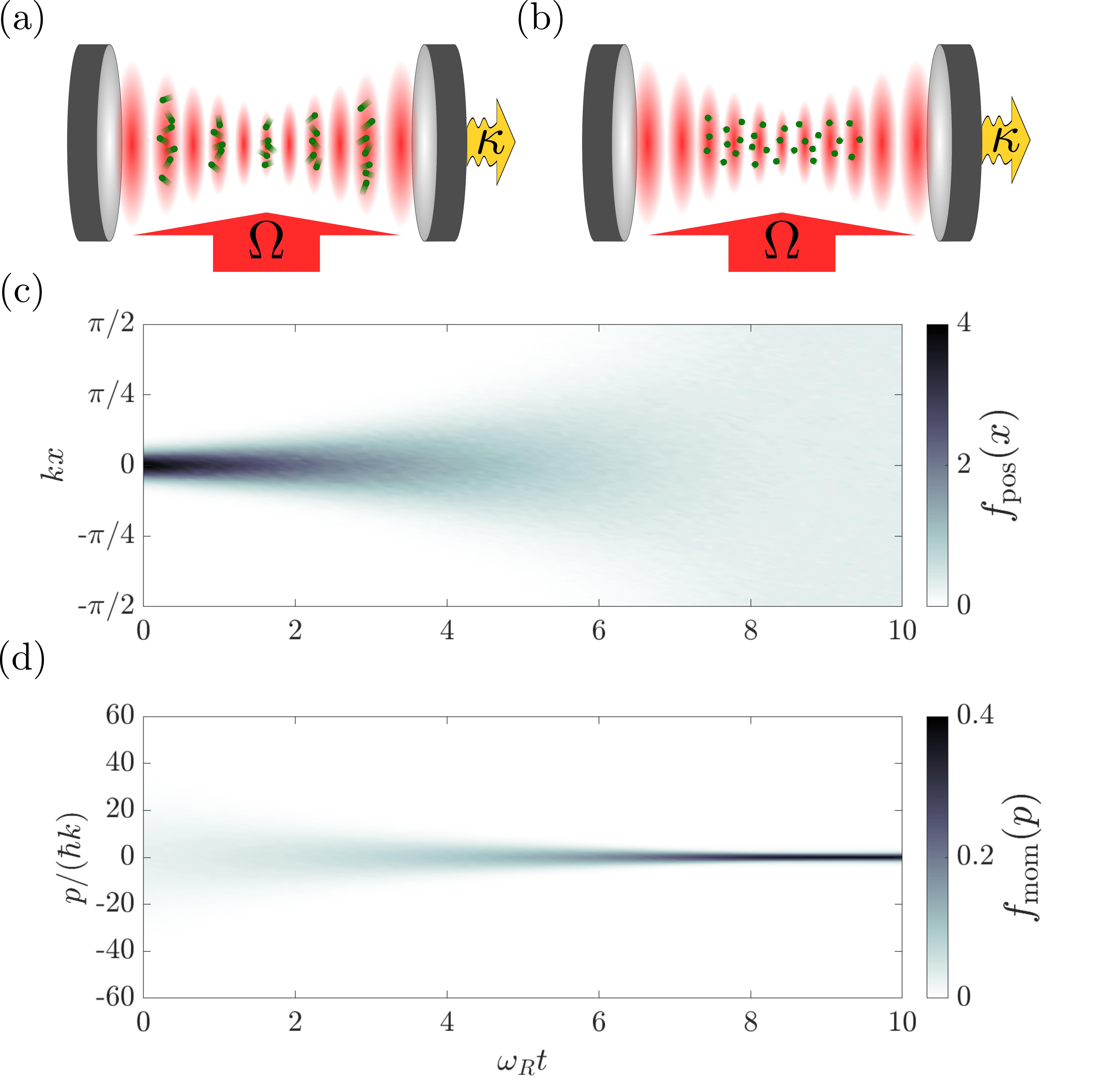}
	\caption{The particles are transversally driven by a laser with Rabi frequency $\Omega$ while dissipation of cavity photons is modeled by $\kappa$. (a) State of the particles after the first stage: the kinetic energy is determined by the cavity linewidth and the particles form a strongly self-organized pattern. (b) State of the particles after the second stage: the kinetic energy is smaller while they are distributed homogeneously in space. The position distribution $f_{\mathrm{pos}}(x)$ (c) as function of $x$ in units of $k^{-1}$ and the momentum distribution $f_{\mathrm{mom}}(p)$ (d) as function of $p$ in units of $\hbar k$ and as functions of time $t$ in units of $\omega_R^{-1}$ in the second stage. The magnetization, determined by the localization of the particles around $kx\approx0$, decreases adiabatically while the kinetic energy decreases as well.\label{Fig:1}}
\end{figure} 

The goal of this work is to present a protocol which can lower the kinetic energy of polarizable particles close to the recoil limit even if the cavity linewidth is orders of magnitude wider. For this, we present a two-stage cooling protocol that uses both, cavity cooling and self-organization to bring the particles to a final kinetic energy that is of the order of the recoil energy. The first stage uses collective cavity cooling of many particles with a high laser power. The final temperature of the particles is here mostly determined by the cavity linewidth while the particles form a strongly self-organized (magnetized) pattern [see Fig.~\ref{Fig:1}(a)]. For these parameters, while the atoms posses a rather high kinetic energy, they are tightly confined in space in a pattern which supports constructive interference of scattered laser photons. In a second stage the laser power is decreased slowly such that the magnetization of the particles is adiabatically decreased [see Fig.~\ref{Fig:1}(c)]. Reminiscent of the magnetocaloric effect and the principle of adiabatic demagnetization~\cite{Tishin:2003}, this results in a decrease of the magnetization of the particles and simultaneously lowers their kinetic energy [see Fig.~\ref{Fig:1}(d)]. In contrast, however, we do not ramp an external magnetic field but the laser driving amplitude which effectively reduces the particle-particle interactions. This principle is also related to so-called release-retrap or adiabatic trap relaxation protocols which are common in optical lattices and used to achieve low temperatures and high phase-space densities~\cite{DePue:1999,Hu:2017}. In such protocols the particles are cooled in tightly confined trapping potentials while a subsequent adiabatic trap relaxation lowers the kinetic energy even further. In contrast, however, the demagnetization presented in this paper is of collective nature and comes from strong cavity-mediated atom-atom interactions instead of deep laser trapping. This is important since it allows us to perform the demagnetization fast enough such that cavity shot noise does not significantly heat the system while decreasing the driving-laser power.  At the end of this ramp, particles reach a final temperature that can be orders of magnitude lower than the one of conventional cavity cooling while the particles reach a spatially homogeneous state [see Fig.~\ref{Fig:1}(b)]. 

The paper is structured as follows. In Sec.~\ref{section:2} we introduce the semiclassical equations that are used to simulate the system. Furthermore we show analytical predictions for the final kinetic energy following an adiabatic ramp. After that in Sec.~\ref{section:3} we analyze the effects of dissipation and show the actual proposed cooling protocol. The conclusions are drawn in Sec.~\ref{section:4} and Appendix~\ref{App:A} provides details of the calculations in Sec.~\ref{section:2}.

\section{Physical setup} \label{section:2}
We consider a setup of $N$ transversally driven polarizable particles with mass $m$ inside a single-mode cavity. The particles are driven far off-resonant with detuning $\Delta_a=\omega_L-\omega_a$ between laser frequency $\omega_L$ and transition frequency $\omega_a$ such that spontaneous emission and the population of the excited state can be neglected. The laser light is thus coherently scattered with rate $S=g\Omega/\Delta_a$ by the particles into the cavity. Here, $\Omega$ is the Rabi frequency of the driving-laser field and $g$ is the vacuum Rabi frequency of the cavity. We assume that the coupling between the particles and the cavity is proportional to the mode function $\cos(kx)$ where $k$ denotes the wave number of the cavity mode. The laser frequency is red-detuned to the frequency $\omega_c$ of the single resonator mode with detuning $\Delta_c=\omega_L-\omega_c<0$. Furthermore the cavity mode looses photons with rate $\kappa$. In what follows we discard effects of the dynamical stark shift $U=g^2/\Delta_a$. This is possible if $\Delta_c$ and $\kappa$ are much larger than $NU$. 
\subsection{Semiclassical description}
We present now a semiclassical description of the particles' center of mass motion and the cavity field. The coupled equations for the particles' motion with position $x_j$ and momentum $p_j$ and the real and imaginary part of the cavity field $\mathcal{E}_{r}$ and $\mathcal{E}_{i}$ evolve according to the following stochastic differential equations \cite{Domokos:2001}
\begin{subequations}
	\begin{align}
		&dx_j=\frac{p_j}{m}dt\,,\label{xwcav}\\
		&dp_j=2\hbar k S \mathcal{E}_{r}\sin(kx_j)dt\,,\label{pwcav}\\
		&d\mathcal{E}_{r}=(-\Delta_c\mathcal{E}_{i}-\kappa \mathcal{E}_{r})dt+d\xi_{r}\,,\label{Er}\\
		&d\mathcal{E}_{i}=(\Delta_c\mathcal{E}_{r}-\kappa \mathcal{E}_{i}-NS \Theta)dt+d\xi_{i}\,,\label{Ei}
	\end{align}
	\label{com+cavity}
\end{subequations}
and $j=1,2,\dots,N$. The noise terms $d\xi_{i},d\xi_{r}$ have vanishing first moments, $\langle d\xi_{i}\rangle=0=\langle d\xi_{r}\rangle$, while the second moments fulfill $\langle d\xi_{i}d\xi_{i}\rangle=\kappa dt/2$, $\langle d\xi_{r}d\xi_{r}\rangle=\kappa dt/2$, and $\langle d\xi_{r}d\xi_{i}\rangle=0$. Further we have defined the order parameter or magnetization $\Theta$ by
\begin{align}
	\Theta=\frac{1}{N}\sum_{j=1}^N\cos(kx_j).\label{Theta}
\end{align}Equation~\eqref{com+cavity} describes the driven-dissipative dynamics of the particles that couple to a dissipative cavity mode. 

To obtain a better understanding of the forces that are mediated by the cavity it is useful to eliminate the cavity degrees of freedom from the dynamics. In this paper we work in the limit where $\kappa,|\Delta_c|\gg k\Delta p/m$ and $\kappa^2\gg\omega_R\sqrt{N}S$, with $\omega_R=\hbar k^2/(2m)$ the recoil frequency. This implies that the cavity degrees evolve much faster and can be adiabatically eliminated~\cite{Schuetz:2013}. Here, $k\Delta p/m$ is the Doppler-width and $\Delta p$ is the single-particle momentum width. Working in this regime allows us to simplify Eq.~\eqref{com+cavity} by calculating the adiabatic stationary state of the cavity field. This is done by formal integration of the differential equations for $\mathcal{E}_r$ and $\mathcal{E}_i$. The adiabatic solution is given by 
\begin{align}
	\mathcal{E}_r=&\frac{\Delta_cNS\Theta}{\Delta_c^2+\kappa^2},\\
	\mathcal{E}_i=&\frac{-\kappa NS\Theta}{\Delta_c^2+\kappa^2}.
	\end{align}
	Using this result in Eqs.~\eqref{com+cavity} results in
\begin{subequations}
	\begin{align}
		&dx_j=\frac{p_j}{m}dt\,,\label{x}\\
		&dp_j=-2kV\sin(kx_j)\Theta dt\,,\label{p}
	\end{align}
	\label{com}
\end{subequations}
with
\begin{align}
	V=-\hbar\Delta_c\frac{NS^2}{\Delta_c^2+\kappa^2}.\label{V}
\end{align}
We emphasize that $V$ is positive since we assumed $\Delta_c<0$ which allows for self-organization and cavity cooling \cite{Asboth:2005}. The dynamics given by Eq.~\eqref{com} can be rewritten using an effective Hamiltonian
\begin{align}
	H_{\mathrm{eff}}=\sum_j\frac{p_j^2}{2m}-NV\Theta^2\,,
\end{align}
with $dx_j/dt=\partial H_{\mathrm{eff}}/\partial p_j,dp_j/dt=-\partial H_{\mathrm{eff}}/\partial x_j$. The term $\propto V$ is a long-range interaction potential which tries to maximize the value of $\Theta$. The latter is the order parameter or magnetization and used to distinguish between the self-organized and the spatially homogeneous phase. In this context the values of $\cos(kx_j)$ can be seen as a continuous magnetization for each atom which takes values between $-1$ and $+1$. In the spatially homogeneous or paramagnetic phase $\cos(kx_j)$ takes random values between $-1$ and $+1$ such that $\Theta\approx 0$. In the self-organized or ferromagnetic phase the particles form a pattern with a periodicity that is determined by the wavelength $\lambda=2\pi/k$ such that $|\Theta|>0$ meaning the individual spins fulfil either all $\cos(kx_j)\approx 1$ or all $\cos(kx_j)\approx -1$. In an experiment the magnetization can be detected from the cavity output. This can be seen by finding the stationary state of Eqs.~\eqref{Er}-\eqref{Ei} that can be used to calculate the intra-cavity photon number
\begin{align}
	I = \left\langle \mathcal{E}_r^2+\mathcal{E}_i^2\right\rangle\approx\frac{N^2S^2}{\Delta_c^2+\kappa^2}\langle\Theta^2\rangle,\label{I}
\end{align}
where the average runs over different initializations and trajectories.

After adiabatic elimination of the cavity degrees of freedom we have derived a dynamical description from a classical Hamiltonian. This implies that Eqs.~\eqref{com} conserve the mean energy $\langle H_{\mathrm{eff}}\rangle$ for a time independent interaction strength $V$. The description by means of Hamiltonian dynamics is, however, only true on a certain timescale where dissipative effects can be discarded \cite{Schuetz:2013,Schuetz:2014,Schuetz:2015,Jaeger:2016,Schuetz:2016}.

In the following we are interested in changing $V$ very slowly such that the particles evolve mainly adiabatically but sufficiently fast such that dissipative effects are negligible. 
\subsection{Adiabatic ramp of the interaction strength}
We assume that the distribution function of the particles is given by a thermal state which can be seen as the stationary state of the system reached after sufficiently long times. This state is given by
\begin{align}
	f_t({\bf x},{\bf p})=Z^{-1}(\beta_t)e^{-\beta_t H_{\mathrm{eff}}}\label{thermalstate}
\end{align}
with single-particle kinetic energy \begin{align}
	E^{\mathrm{kin}}(t)=\frac{\langle p^2\rangle(t)}{2m}=\frac{1}{2\beta_t},\label{Ekin}
\end{align}
and partition function $Z(\beta_t)=\int d{\bf x}\int d{\bf p}\,e^{-\beta_t H_{\mathrm{eff}}}$. Note that $f_t$ and $\beta_t$ are explicitly time dependent.
The expectation value is here defined by $\langle h({\bf x},{\bf p})\rangle(t)=\int d{\bf x}\int d{\bf p}\,h({\bf x},{\bf p})f_t({\bf x},{\bf p})$ with integrals $\int d{\bf x}=\int_{0}^{\lambda}dx_1...\int_{0}^{\lambda}dx_N$ and $\int d{\bf p}=\int_{-\infty}^{\infty}dp_1...\int_{-\infty}^{\infty}dp_N$ and for an arbitrary function $h({\bf x},{\bf p})$ of the atomic positions and momenta.

Now, we assume a time dependent $V$ and in particular we assume that the temporal change of $V$ is sufficiently slow such that the particles remain in a thermal state. With this assumption we are able to derive a dynamical equation for the kinetic energy in the following.

Using Eqs.~\eqref{com} and Eq.~\eqref{Ekin} we obtain the dynamical evolution of the single-particle kinetic energy
\begin{align}
	\frac{dE^{\mathrm{kin}}}{dt}=V\frac{d\langle \Theta^2\rangle}{dt}. \label{Ekindyn0}
\end{align}
We may further write
\begin{align*}
	\langle \Theta^2\rangle=\left.\frac{dF}{dy}\right|_{y=\frac{NV}{2E^{\mathrm{kin}}}}
\end{align*}
with
\begin{align}
	F(y)=\ln\left(\int d{\bf x}\,e^{y \Theta^2}\right).\label{F}
\end{align}
Then, using $V=2E^{\mathrm{kin}}y/N$ we can rewrite Eq.~\eqref{Ekindyn0} as
\begin{align}
	N\frac{dE^{\mathrm{kin}}}{dy}=2E^{\mathrm{kin}}y\frac{d^2F(y)}{dy^2}, \label{Ekindyn1}
\end{align}            
and integration of this equation leads to
\begin{align}
	\int_{E^{\mathrm{kin}}_0}^{E^{\mathrm{kin}}_1}\frac{dE^{\mathrm{kin}}}{E^{\mathrm{kin}}}=\int_{y_0}^{y_1} dy\frac{2y}{N}\frac{d^2F}{dy^2},
\end{align}
where $y_n=NV_n/(2E^\mathrm{kin}_n)$ with $n=0,1$.
The latter can be solved using integration by parts and we obtain
\begin{align}
	\ln\left(\frac{E^{\mathrm{kin}}_1}{E^{\mathrm{kin}}_0}\right)=\frac{2}{N}\left[y\frac{dF(y)}{dy}-F(y)\right]_{y_0}^{y_1},\label{onestepbefore}
\end{align}
where we used the notation $[f(y)]_{y_0}^{y_1}=f(y_1)-f(y_0)$.
Now defining $\alpha_n=y_n/N=V_n/(2E^\mathrm{kin}_n)$ and performing the limit $N\to\infty$ with $\alpha_n=\mathrm{const}$. we obtain the result
\begin{align}
	\frac{E^{\mathrm{kin}}_1}{E^{\mathrm{kin}}_{0}}=\left[\frac{I_0(2\alpha_{0}\theta(\alpha_{0}))e^{-2\alpha_{0}\theta^2(\alpha_{0})}}{I_0(2\alpha_1\theta(\alpha_1))e^{-2\alpha_1\theta^2(\alpha_1)}}\right]^{2},\label{explicitresult}
\end{align}
where $I_n$ is the $n$th modified Bessel function and $\theta(\alpha)$ describes the stable solution of the equation
\begin{align}
	\theta=\frac{I_1(2\alpha \theta)}{I_0(2\alpha \theta)}.\label{fixpointequation}
\end{align}
For a detailed derivation see Appendix \ref{App:A}. The value for $\theta$ calculated from Eq.~\eqref{fixpointequation} is the mean magnetization of the particles for the given value of $\alpha$. Thus Eq.~\eqref{explicitresult} connects the magnetization before and after the ramp with the kinetic energy before and after the ramp. We now discuss how this result can be used to lower the kinetic energy of the particles.

\begin{figure}[h!]
	\center
	\includegraphics[width=1\linewidth]{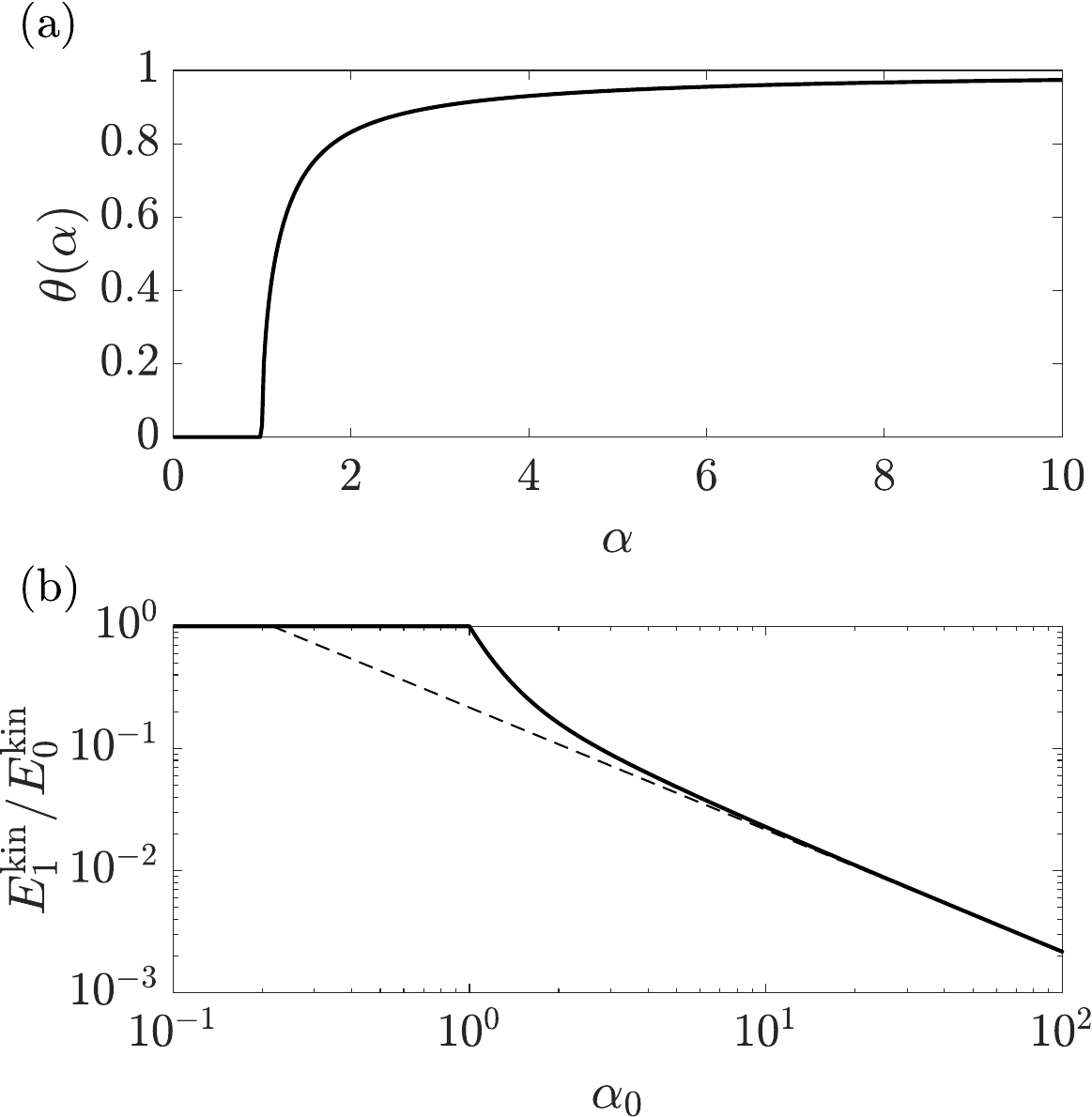}
	\caption{(a) The stable solution $\theta(\alpha)$ of Eq.~\eqref{fixpointequation} as function of $\alpha$. (b) The quotient of final $E_1^{\mathrm{kin}}$ and initial kinetic energy $E_0^{\mathrm{kin}}$ depending on $\alpha_0$ given by Eq.~\eqref{Eramp}. The dashed line is the asymptotic result given by Eq.~\eqref{asymptoticresult}.\label{Fig:2}}
\end{figure}

In Fig.~\ref{Fig:2}(a) we plotted $\theta(\alpha)$ as a function of $\alpha$. We see that $\theta(\alpha)$ is zero for $\alpha<1$ (paramagnetic phase) and increases for $\alpha>1$ (ferromagnetic phase) while it tends to $1$ in the case $\alpha\to\infty$. Note that we have only shown here the positive solution $\theta>0$ but there is also the solution $-\theta$. This transition from spatially homogeneous (paramagnetic) to self-organized (ferromagnetic) has been described as a phase transition.

In the spatially homogeneous phase, for $\alpha_1,\alpha_0\leq1$, the quotient of the kinetic energies in Eq.~\eqref{explicitresult} is always one. This implies that any adiabatic change within the spatially homogeneous region will to good approximation not affect the kinetic energy.

However, when we assume that the coupling strength is initialized such that the particles are in the self-organized phase and ramped to a value for that the particles are distributed spatially homogeneous, i.e., $\alpha_0>1$ and $\alpha_1\leq 1$, we obtain 
\begin{align}
	\frac{E^{\mathrm{kin}}_1}{E^{\mathrm{kin}}_{0}}=\left[I_0(2\alpha_{0}\theta(\alpha_{0}))e^{-2\alpha_{0}\theta^2(\alpha_{0})}\right]^{2}.\label{Eramp}
\end{align}
We show this result of the right-hand side of  Eq.~\eqref{Eramp} in Fig.~\ref{Fig:2}(b) as black solid line. It is a monotonous decreasing function with $\alpha_0$. Therefore, we conclude that a potentially very low kinetic energy can be reached by starting the ramp from a high coupling strength $V_0$. In this regime, for $\alpha_0\gg1$, we get the asymptotic result
\begin{align}
	\frac{E^{\mathrm{kin}}_1}{E^{\mathrm{kin}}_{0}}=\frac{e}{4\pi \alpha_0},\label{asymptoticresult}
\end{align}
where $e$ is the Euler number.
This shows that the ratio of the kinetic energies is proportional to $1/\alpha_0$ and is plotted as dashed gray line in Fig.~\ref{Fig:2}(b). 

We now discuss how this principle might be applicable to the driven-dissipative dynamics of particles in a cavity.
\section{Cooling protocol} \label{section:3}
In order to apply the results of the previous section we first need to analyze the dissipative effects in the particles' dynamics. This is done by comparing the results of Eq.~\eqref{com}, that discard any dissipative effects with the dynamics including dissipation, Eq.~\eqref{com+cavity}. 
\subsection{Effects of Dissipation}
\begin{figure}[h!]
	\center
	\includegraphics[width=1\linewidth]{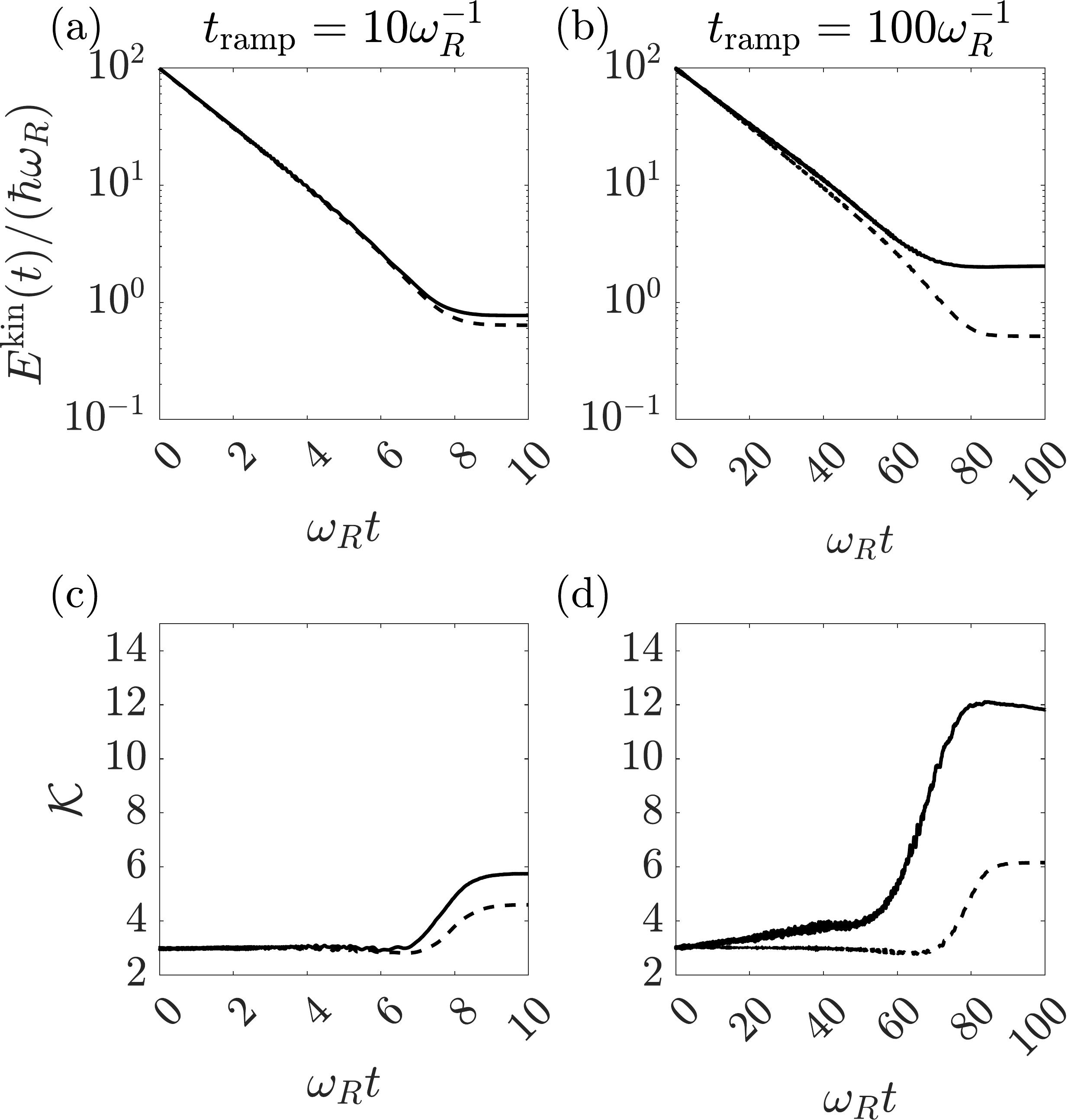}
	\caption{The kinetic energies $E^{\mathrm{kin}}$ in units of $\hbar \omega_R$ (a), (b) and the Kurtosis $\mathcal{K}$ (Eq.~\eqref{Kurtosis}) (c), (d) as function of time $t$ in units of $\omega_R^{-1}$ following a ramp according to Eq.~\eqref{ramp} with $t_{\mathrm{ramp}}=10\omega_R^{-1}$ and $t_{\mathrm{ramp}}=100\omega_R^{-1}$, respectively. We have chosen $V_0=10^4\hbar\omega_R$, $\Delta_c=-\kappa$, $\kappa=400\omega_R$, $N=100$, $NS_0^2=50\kappa^2$ and averaged over $200$ trajectories. The system is initialized in a strongly self-organized thermal state with $E^{\mathrm{kin}}(0)=100 \hbar \omega_R$.\label{Fig:3}}
\end{figure}

We first initialize the particles in a strongly self-organized thermal state with kinetic energy $E^{\mathrm{kin}}$. Then we ramp $V$ according to an exponential ramp 
\begin{align}
	V(t)=V_0\cdot 10^{-5\frac{t}{t_{\mathrm{ramp}}}},\label{ramp}
\end{align}
for different ramping times $t_{\mathrm{ramp}}$. While the choice of an exponential ramp is a technical detail it allows for a rather fast change of the interaction strength for large values of $V\approx V_0$ and slow changes for $V\gtrsim0$. We believe that this is a good compromise between being fast and remaining approximately adiabatic. We study the dynamics of the full system including the cavity degrees of freedom [Eq.~\eqref{com+cavity}] and the dynamics where the cavity degrees of freedom are eliminated [Eq.~\eqref{com}]. Figure~\ref{Fig:3}(a) and Fig.~\ref{Fig:3}(b) show the dynamics of the kinetic energy following the ramp for a ramping time of $t_{\mathrm{ramp}}=10\omega_R^{-1}$ and $t_{\mathrm{ramp}}=100\omega_R^{-1}$, respectively. To put these values into actual numbers we provide an explicit example and use the value $\omega_R=2\pi\times 2\text{ kHz}$ from Ref.~\cite{Hosseini:2017} for $^{133}$Cs which results in ramping times $t_{\mathrm{ramp}}\approx1\text{ ms}$ and $t_{\mathrm{ramp}}\approx10\text{ ms}$, respectively. The dashed line shows the result using the conservative dynamics [Eq.~\eqref{com}] while the solid line represents the full dissipative dynamics [Eq.~\eqref{com+cavity}]. Both curves for both ramping times show a decrease of the kinetic energy. We find a good agreement of both dynamics on short timescales while we observe discrepancies for the longer ramping time. Therefore, we expect that for sufficiently short times dissipative effects are still negligible while they affect the dynamics on longer timescales. This relies on a timescale separation of dissipative and conservative forces that relies on (i) the number of particles and (ii) the typical timescale separation of motion and cavity relaxation, i.e $k\Delta p/m\ll \kappa$. This has been studied in Refs.~\cite{Schuetz:2016,Jaeger:2016} and observed in Ref.~\cite{Wu:2023}.  In conclusion, this preliminary analysis demonstrates that there must be an optimal ramping time for which 
we can achieve the lowest possible temperature.

While our original assumption was that the ramp is close to adiabatic we expect this assumption to fail, especially because we ramp the system parameters across a phase transition. An observable to test this is the Kurtosis 
\begin{align}
	\mathcal{K}(t)=\frac{\langle p^4\rangle(t)}{[\langle p^2\rangle(t)]^2}.\label{Kurtosis}
\end{align}
The Kurtosis is $\mathcal{K}=3$ for a Gaussian state and deviates from $3$ for non-Gaussian states. Figure~\ref{Fig:3}(c) and Fig.~\ref{Fig:3}(d) show the Kurtosis with the same labeling for the two different ramping times. In Fig.~\ref{Fig:3}(c) we observe that the Kurtosis remains close to $3$ for times $t\lesssim 7\omega_R^{-1}$. While it deviates for longer times as soon as the value of $V$ crosses the phase transition line. In Fig.~\ref{Fig:3}(d) we observe the same for the simulations of the conservative dynamics [Eq.~\eqref{com}] and times $t\lesssim 70\omega_R^{-1}$ while the simulation of the full dissipative dynamics [Eq.~\eqref{com+cavity}] shows values of $\mathcal{K}\neq 3$ on much shorter timescales. Our finding support our claim that the dynamics does not remain adiabatic across the phase transition. In addition, the discrepancies between the conservative and dissipative dynamics, predict an dissipation-assisted creation of non-Gaussian states that we believe is closely related to the dissipation-assisted stabilization of non-Gaussian states predicted in Ref.~\cite{Schuetz:2016}.

\begin{figure}[h!]
	\center
	\includegraphics[width=1\linewidth]{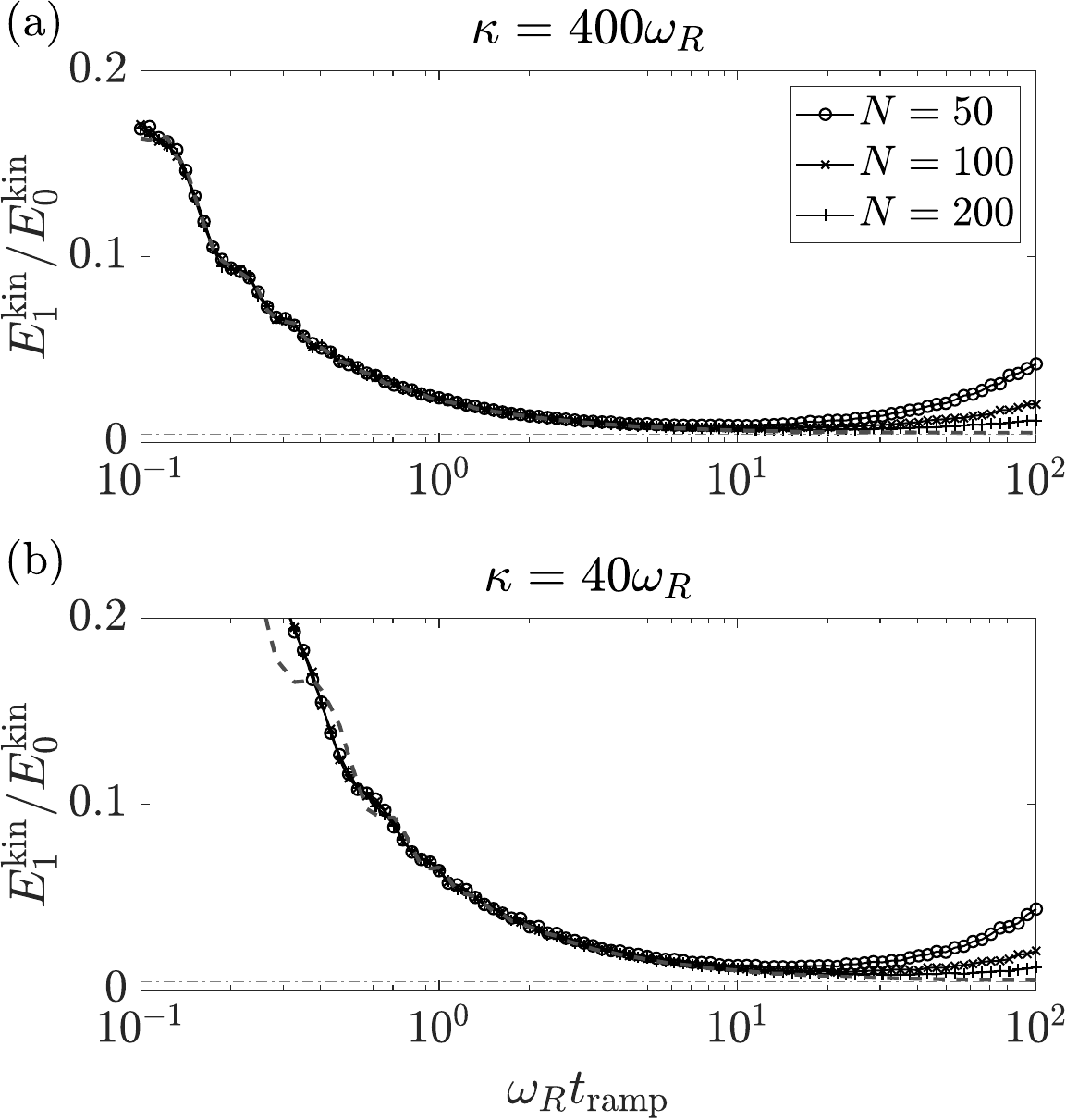}
	\caption{The final kinetic energy $E^{\mathrm{kin}}_{1}=E^{\mathrm{kin}}(t_\mathrm{ramp})$ in units of $E^{\mathrm{kin}}_{0}=\hbar \kappa/4$ for different ramping times $t_{\mathrm{ramp}}$ in units of $\omega_R^{-1}$ and for (a) $\kappa=400\omega_R$ and (b) $\kappa=40\omega_R$. The simulations have been performed for different particle numbers (see inset of (a)) using Eqs.~\eqref{com+cavity}. The dashed gray lines correspond to simulations of Eqs.~\eqref{com} with $N=200$ particles. The horizontal dashed-dotted gray lines are the predictions of Eq.~\eqref{asymptoticresult}. All results are obtained for simulations with parameters $\Delta_c=-\kappa$, $\alpha_0=V_0/(2E_1^{\mathrm{kin}})=50$ and using $20000/N$ trajectories. \label{Fig:4}}
\end{figure}

We now analyze the dependence of the minimum achievable temperature on the ramping time $t_{\mathrm{ramp}}$. 
In Fig.~\ref{Fig:4} we compared the values of the final kinetic Energies $E^{\mathrm{kin}}_{1}=E^{\mathrm{kin}}(t_\mathrm{ramp})$ for different ramping times $t_{\mathrm{ramp}}$, different particle numbers, and different ratios of $\kappa/\omega_R$. The black line with symbols are calculated using simulations of Eqs.~\eqref{com+cavity} with $N=50$ (circles), $N=100$ (crosses), and $N=200$ (pluses) where we show $\kappa=400\omega_R$ in Fig.~\ref{Fig:4}(a) and $\kappa=40\omega_R$ in Fig.~\ref{Fig:4}(b). These are realistic values and the lower value of $\kappa=40\omega_R$ is close to the one realized in Ref.~\cite{Hosseini:2017}. For both simulations we have initialized the system with a pumping strength of $\alpha_0=V_0/(2E^{\mathrm{kin}}_0)=50$ and $E^{\mathrm{kin}}_1=\hbar \kappa/4$. The results of the simulations predict a local minimum of the kinetic energy in the range $\omega_Rt_{\mathrm{ramp}}=10-100$ thus showing that there is an optimal ramping time. In general we observe that this optimal ramping time is shorter for smaller particle numbers. In addition, also the minimum achievable kinetic energy is larger for smaller particle numbers. We expect that this is due the timescale separation between the conservative and dissipative forces that becomes larger for increasing particle numbers. 
We also find that the result of the kinetic energy for $\kappa=40\omega_R$ (b) appear to be slightly displaced to larger ramping times with respect to the simulations for $\kappa=400\omega_R$ (a). We expect that this is due to a violation of the adiabaticity criteria, $\kappa t_{\mathrm{ramp}}\gg1$ and $k\Delta p t_{\mathrm{ramp}}/m\gg1$, that are not fulfilled for smaller values of $k\Delta p/m$ and $\kappa$ and short ramping times in Fig.~\ref{Fig:4}(b). For completeness we have also included a simulation of Eq.~\eqref{com} that does not include dissipation and noise. The results are visible as gray dashed lines in Fig.~\ref{Fig:4}. Those curves are monotonically decreasing thus showing that noise and dissipation are the origins for the local minima in the kinetic energy in the full simulations. The theoretical minimum of the achievable kinetic energy is shown as gray dashed-dotted line. It is calculated using Eq.~\eqref{asymptoticresult} and $\alpha_0=50$ resulting in $E^{\mathrm{kin}}_1/E^{\mathrm{kin}}_0\approx0.04$. We observe that the simulation without dissipation and noise (gray dashed line) is converging to this theoretical minimum in the limit $\omega_Rt_{\mathrm{ramp}}\to\infty$.

\subsection{Cooling protocol}
We now use this gained insight to minimize the kinetic energy of particles that are initially in a spatially homogeneous configuration. Hereby we assume that the initial state is a thermal state with temperature $k_{\mathrm{B}}T_{\mathrm{in}}=\hbar \kappa/2$ that can be reached by cavity cooling for $\Delta_c=-\kappa$. The actual choice of this state is rather arbitrary but should be sufficiently cold such that the ensemble can be cavity (laser) cooled.

In a first stage we perform a quench in the driving-laser intensity determined by $S$ such that $V$ has a value $V_\mathrm{fer}$ for that the system reaches a state well inside the self-organized phase. On a very long time the system again reaches a stationary state which is thermal. To be consistent, for large laser intensities we need to take corrections of the final temperature into account which come from the laser driving power. This final kinetic energy is given in the well-organized regime (see Ref. \cite{Niedenzu:2011,Griesser:2012}) by 
\begin{align}
	E^{\mathrm{kin}}_{\mathrm{fer}}=\frac{k_BT_{\mathrm{fer}}}{2}=\frac{\hbar (\Delta_c^2+\kappa^2+4\omega_0^2)}{-8\Delta_c},\label{T}
\end{align}
where $\omega_0^2=4\omega_R\frac{V_\mathrm{fer}}{\hbar}$ is the effective trapping frequency. 

In a second stage we consider a ramp from $V_\mathrm{fer}$ back to a value close to zero such that both magnetization and kinetic energy are adiabatically reduced. If we assume that the system remains adiabatic in a thermal state the optimum final kinetic energy can be approximated using Eq.~\eqref{asymptoticresult} by
\begin{align}
	E^{\mathrm{kin}}_{\mathrm{par}}=\frac{e}{2\pi}\frac{(E^{\mathrm{kin}}_{\mathrm{fer}})^2}{V_{\mathrm{fer}}}.\label{Epara}
\end{align}
Minimizing Eq.~\eqref{Epara} with respect to $V_{\mathrm{fer}}$ we find 
\begin{align}
	E^{\mathrm{kin}}_{\mathrm{min}}=\frac{e}{2\pi}\hbar\omega_R\frac{\Delta_c^2+\kappa^2}{\Delta_c^2}\label{minkin}
\end{align}
at an optimum value
\begin{align}
	V_\mathrm{fer}^{\mathrm{opt}}=\frac{\hbar(\Delta_c^2+\kappa^2)}{16\omega_R}.\label{Vferopt}
\end{align}
This minimum kinetic energy is of the order of the recoil energy $\hbar \omega_R$.

\begin{figure}[h!]
	\center
	\includegraphics[width=1\linewidth]{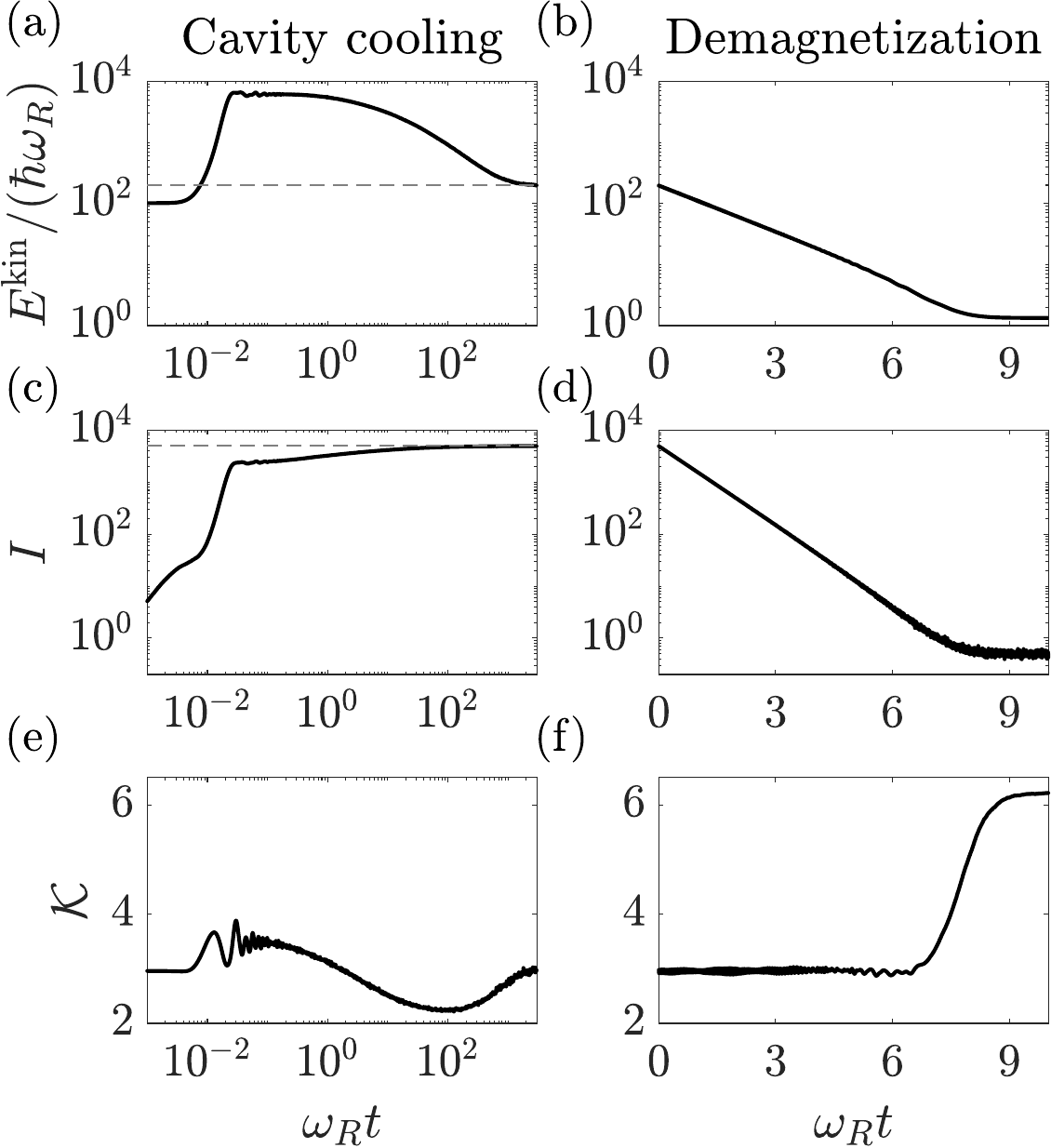}
	\caption{Dynamics of the kinetic energies $E^{\mathrm{kin}}$ in units of $\hbar \omega_R$ (a), (b), the cavity field determined by Eq.~\eqref{I} (c), (d), and the Kurtosis $\mathcal{K}$ [Eq.~\eqref{Kurtosis}] (e), (f) as function of time $t$ in units of $\omega_R^{-1}$. The plots in (a), (c), (e) are obtained after a quench from $V\approx 0$ to $V=V_\mathrm{fer}^{\mathrm{opt}}$ [Eq.~\eqref{Vferopt}] where the particles are initialized in a spatially homogeneous state with Gaussian momentum distribution and an initial kinetic energy $E^{\mathrm{kin}}(0)=\hbar\kappa/4$. After a relaxation time $t_{\mathrm{f}}=3\times10^3\omega_R^{-1}$ we perform a ramp according to Eq.~\eqref{ramp} resulting in the dynamics visible in (b), (d), (f). The ramping time is $t_{\mathrm{ramp}}=10\omega_R^{-1}$ and all simulations have been performed with $\Delta_c=-\kappa$, $\kappa=400\omega_R$, $N=100$, and averaging over $200$ trajectories. \label{Fig:5}}
\end{figure}

Now, we show simulations that tests this prediction. Following the procedure of the first stage, we show in Fig.~\ref{Fig:5}(a) the dynamics of the kinetic energy after a quench from $V\approx 0$ to $V=V_\mathrm{fer}^{\mathrm{opt}}$ [Eq.~\eqref{Vferopt}]. Initially we observe a fast increase in the magnetization determined by $\left\langle\Theta^2\right\rangle$. This can be seen in Fig.~\ref{Fig:5}(c) where plot the cavity field determined by Eq.~\eqref{I}. Since on short timescales the energy is conserved the kinetic energy is also exponentially increasing reaching a maximum of $E^{\mathrm{kin}}\approx6.5\times 10^3\,\hbar\omega_R$. On longer timescales dissipation guides the system towards a thermal state with a temperature given by Eq.~\eqref{T} (horizontal gray dashed line in Fig.~\ref{Fig:5}(a)) with the corresponding magnetization (horizontal gray dashed line in Fig.~\ref{Fig:5}(c)). The steady-state magnetization has been calculated using $\langle\Theta^2\rangle\approx\theta^2$ where $\theta$ is the solution of Eq.~\eqref{fixpointequation}.  In Fig.~\ref{Fig:5}(e) we have plotted the Kurtosis (Eq.~\eqref{Kurtosis}) that starts and ends at a value close to $\mathcal{K}\approx3$ that suggests that both, the final state and the initial state are thermal. 

Now, following the second stage, we ramp the coupling strength according to Eq.~\eqref{ramp} with a ramping time $t_{\mathrm{ramp}}=10\omega_R^{-1}$ that we have found to be close to optimal for our choice of parameters in the previous subsection. Figure~\ref{Fig:5}(b) shows the decrease of the kinetic energy that eventually reaches a value that is of the order of the recoil energy. We observe a final kinetic energy of $E^{\mathrm{kin}}_{\mathrm{final}}=1.3\hbar \omega_R$ whereby Eq.~\eqref{minkin} predicts a similar value of $E^{\mathrm{kin}}_{\mathrm{min}}\approx0.9\hbar\omega_R$. During this process the field intensity and the mean magnetization of the system is decreased as visible in Fig.~\ref{Fig:5}(d) reminiscent of adiabatic demagnetization. We emphasize that the adiabatic process is much faster than cavity cooling in the first stage. This is visible by comparing the time axes in Fig.~\ref{Fig:5}(a) and Fig.~\ref{Fig:5}(b). This difference in the timescales comes from the fact that the adiabatic process is collective. The Kurtosis, visible in Fig.~\ref{Fig:5}(f), remains 3 during the ramp until the phase transition line is crossed and the process is not adiabatic anymore. 

\section{Conclusion}\label{section:4}

In this paper we have studied the possibility to cool transversally driven particles inside of an optical cavity using a combination of cavity cooling and a protocol reminiscent of adiabatic demagnetization. To analyze the effect of dissipation we have performed simulations of dissipative and conservative dynamical models for this physical setup. We have shown that the particles can reach kinetic energies comparable to the recoil limit that is for our parameter choice below the typical limit of cavity cooling. To achieve this final kinetic energy we have tuned the laser power from a value well above the self-organization threshold to below this threshold. The duration of this ramp is chosen sufficiently long such that it seems to be quasi-adiabatic for the coherent dynamics but sufficiently fast such that dissipation has only a minor effect on the final kinetic energy. 

While the results presented here rely on adiabatically changing the coupling or interaction strength that results in a change of the internal magnetization, we expect that one can achieve similar physics by changing an additional external field that simulates an effective magnetic field. This can for instance be accomplished by modulating a laser that is directly driving the cavity beside the transversal laser field~\cite{Niedenzu:2013}.  

We want to emphasize that the cooling stage is crucial to achieve the final kinetic energy as it reduces the entropy. However, the precise cooling protocol is rather arbitrary. In particular, we want to emphasize that the choice of cavity cooling, visible in Fig.~\eqref{Fig:5}, can be replaced by other schemes, e.g. a ramp of the interaction strength instead of a quench or even by other laser cooling mechanisms. Important is solely that the outcome of this first stage is the preparation of a sufficiently cold and highly self-organized (magnetized) particle ensemble. 

Regarding the ultimate limits of this cooling protocol we have to remark that our analysis is performed with semiclassical equations. This means our approach is only valid for kinetic energies that are above the recoil limit. In addition, we have not included the quantum statistics of the particles which becomes relevant for low temperatures. We believe that including the latter would be an interesting extension of our work since one might expect different distributions for Bosons~\cite{Baumann:2010} and Fermions~\cite{Zhang:2021,Helson:2023}. In future work, it might also be interesting to use multi-mode cavities that provide more possibilities to tune the interactions and dissipation \cite{Torggler:2014,Keller:2017,Keller:2018}. In general we believe that the study of such systems is not only interesting for advances in laser and cavity cooling but is also interesting as a simulator for classical and quantum thermodynamics \cite{Vinjanampathy:2016,Niedenzu:2018}. In conclusion, engineering interactions and dissipation for particles in optical cavities is a versatile tool for quantum technologies and to study new physics.  
\begin{acknowledgements}
SBJ acknowledges stimulating discussions with Stefan Sch\"utz, John Cooper, and Giovanna Morigi. We acknowledge support from the Deutsche Forschungsgemeinschaft (DFG, German Research Foundation) under project number 277625399 - TRR 185 (B4) and under Germany’s Excellence Strategy – Cluster of Excellence Matter and Light for Quantum Computing (ML4Q) EXC 2004/1 – 390534769.

\end{acknowledgements} 
	\bibliography{article.bib} 
	\appendix
	\section{Derivation of the mean magnetization} \label{App:A}      
	In this appendix we describe how we obtain an analytical expression for the quantity $\exp[F(y)]$ defined in Eq.~\eqref{F} in the $N\to\infty$ limit. 
	
	We use the Hubbard-Stratonovich transformation to obtain
	\begin{align*}
		e^{y\Theta^2}=\sqrt{\frac{y}{\pi}}\int_{-\infty}^{\infty} d\theta\,e^{-y \theta^2+2y\theta \Theta}.
	\end{align*}
	Integrating leads to
	\begin{align*}
		\int d{\bf x}\,e^{y\Theta^2}=\lambda^N\sqrt{\frac{y}{\pi}}\int_{-\infty}^{\infty} d\theta\,e^{-y \theta^2+N\ln\left(I_0(2\theta y/N)\right)}.
	\end{align*}
	Defining now $\alpha=y/N$ we obtain
	\begin{align*}
		\int d{\bf x}\,e^{N\alpha\Theta^2}=\lambda^N\sqrt{\frac{N\alpha}{\pi}}\int_{-\infty}^{\infty} d\theta\,e^{-N\mathcal{F}(\theta)},
	\end{align*}
	with the single-particle free energy
	\begin{align}
		\mathcal{F}(\theta)=\alpha\theta^2-\ln(I_0(2\alpha\theta)).\label{freeenergy}
	\end{align}
	In the large $N$ limit we can use a saddlepoint approximation and find
	\begin{align}
		\int d{\bf x}\,e^{N\alpha\Theta^2}=\lambda^N\sqrt{\frac{2\alpha}{H(\theta_0)}}e^{-N\mathcal{F}(\theta_0)}.\label{relation}
	\end{align}           
	where $\theta_0$ is the minimum of $\mathcal{F}(\theta_0)$. The quantity ${H(\theta_0)=d^2\mathcal{F}(\theta_0)/(d\theta^2)}$ is the second derivative with respect to $\theta$ evaluated at the point $\theta_0$.  In order to find the minimum of the free energy we use $d\mathcal{F}(\theta_0)/(d\theta)=0$ that results in the fixpoint relation
	\begin{align}
		\theta_0=\frac{I_1(2\alpha\theta_0)}{I_0(2\alpha\theta_0)}.\label{fixpoint}
	\end{align}
	We call a solution of Eq.~\eqref{fixpoint} that is a minimum of $\mathcal{F}$, Eq.~\eqref{freeenergy} a stable solution.
	
	To simplify Eq.~\eqref{onestepbefore} we can now use
	\begin{align}
		\frac{2}{N}y\frac{d}{dy}F(y)=\frac{2y}{N}\frac{\int d{\bf x}\Theta ^2e^{y\Theta^2}}{\int d{\bf x}e^{y\Theta^2}}\approx2\alpha\theta(\alpha)^2\label{eq:1}
	\end{align}
	and
	\begin{align}
		-\frac{2}{N}F(y)\approx2\mathcal{F}(\alpha)=2\alpha\theta^2(\alpha)-2\ln(I_0(2\alpha\theta))\label{eq:2}
	\end{align}
	with $\alpha=y/N$. To obtain Eq.~\eqref{explicitresult} we can then simply use
	\begin{align}
	\frac{E^{\mathrm{kin}}_1}{E^{\mathrm{kin}}_0}=\frac{e^{\frac{2}{N}y_1F'(y_1)-\frac{2}{N}F(y_1)}}{e^{\frac{2}{N}y_01F'(y_0)-\frac{2}{N}F(y_0)}}
	\end{align}
	and the results in Eq.~\eqref{eq:1} and Eq.~\eqref{eq:2}.
\end{document}